# Energy Efficient Software Matching in Distributed Vehicular Fog Based Architecture with Cloud and Fixed Fog Nodes

Rui Ma, Amal A. Alahmadi, Taisir E. H. El-Gorashi, and Jaafar M. H. Elmirghani
*School of Electronic & Electrical Engineering, University of Leeds, LS2 9JT, United Kingdom*
e-mail: {ml16r5m, elaaal, t.e.h.elgorashi, j.m.h.elmirghani}@leeds.ac.uk

**ABSTRACT**
The rapid development of vehicles on-board units and the proliferation of autonomous vehicles in modern cities create a potential for a new fog computing paradigm, referred to as vehicular fog computing (VFC). In this paper, we propose an architecture that integrates a vehicular fog (VF) composed of vehicles clustered in a parking lot with a fixed fog node at the access network and the central cloud. We investigate the problem of energy efficient software matching in the VF considering different approaches to deploy software packages in vehicles.

## 1. INTRODUCTION

The increase in the number of end-users and data intensive applications has led to tremendous growth in Internet traffic and an increase in the demands placed on data centers and cloud services which resulted in an increase in the power consumption of networks [1]. Given the environmental and economic impact, recent research efforts proposed energy-efficient solutions to reduce networks power consumption [2]-[5] through network virtualization [6]-[8], optimal design of network architecture [9]-[13], content distribution [14]-[16], renewable energy [17], big data networks [18]-[21] and network coding [22], [23].

Fog computing where computing resources at the network edge are used to reduce the burden on central clouds is also considered to improve the energy efficiency [24]-[26]. The large number of vehicles equipped with advanced embedded on-board units in the modern city create a potential for a new fog computing paradigm, referred to as vehicular fog computing (VFC) [27]. These smart vehicles can form a cluster sharing their on-board resources to form a vehicular fog (VF) node at the network edge that can locally process end-users data [28]. Such architecture is recognized as an energy-efficient paradigm compared to the conventional centralized clouds [29], [30]. Fog architecture has the capability to serve the time-intensive requests and reduce power consumption since the fog nodes located in the network edge shorten the distance between users and servers [31].

Recent research efforts in VF are focused on traffic management in the modern city [32], [33], resource allocation in VFs [34], [35], deployment and dimensioning problem [36], privacy and security issue [37], quality of service [38], the feasibility of VFs and delay minimization [39], [40] and energy efficiency [29], [30]. Limited by their capacity, the vehicle in a VF may not be equipped with all the software packages required to serve different application requests. The software matching problem in the vehicular fog was considered in our previous work [41]. In this paper, we extend our work in [41], by introducing an architecture that integrates a vehicular fog (VF) made up of vehicles clustered in a parking lot; with a fixed fog node at the access network and the central cloud. We study in this new proposed distributed processing architecture the impact of different software package deployment approaches on the power savings achieved by the VF.

The rest of this paper is organized as follows: the proposed architecture and energy-efficient software matching problem are introduced in Section 2. In Sections 3, the results are presented and analyzed. Finally, Section 4 concludes the paper.

## 2. A CLOUD-FOG-VEHICULAR FOG ARCHITECTURE AND SOFTWARE MATCHING PROBLEM

Figure 1 shows the proposed architecture. It is composed of three layers of computing nodes; a central cloud at the core network, a fixed fog node at the access network and vehicles equipped with on-board units clustered in a parking lot. The VF is connected to the fixed fog node and the central cloud through a passive optical network (PON) and an intermediate IP over WDM core network.

The fixed fog node is attached to the Optical Line Terminal (OLT). Requests that require certain software packages to be processed are generated by the users and sent wirelessly to a roadside unit (RSU). The RSU is aware of the deployment of software packages and the availability of computing resources in the vehicles, the fixed fog node and the central cloud. Based on this knowledge the RSU allocates computing resources to the requests matching each request to the computing node with the required software package. The cloud and the fixed fog node are assumed to have all the software packages while certain number of software packages are preloaded to each vehicle in the VF.

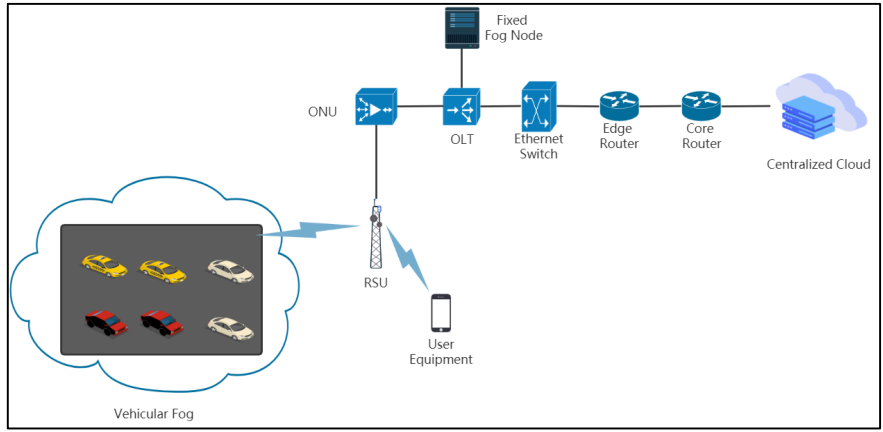

*Figure 1. Vehicles Fog Architecture*

A mixed integer linear programming (MILP) model is developed to optimize processing of requests in the proposed architecture considering the software matching problem so that the total power consumption is minimized. The total power consumption of the proposed architecture is composed of the processing power consumption, networking power consumption and storage power consumption of the central cloud, the fixed fog node, the VF, the wireless network and the optical network.

To evaluate the effect of software package deployment in vehicles on the overall power consumption, we focus on two parameters; the type of software packages and the number of software packages preloaded in each vehicle. We examined two software package deployment approaches. The first approach is the same-type software package deployment where the software packages preloaded in all vehicles are of the same type. The other approach is the random-type software package deployment where software packages are randomly selected from the software packages pool to be downloaded to the vehicles.

### 3. RESULTS AND DISCUSSION
We evaluate the power consumption of the Cloud-Fog-VF architecture considering the two approaches described above and compare it to processing requests in the cloud and the fog. Table I gives the networking, the processing and the storage capacity and power consumption of the different processing layers.

We consider 50 user requests with processing demands, specified in CPU MHz, randomly distributed between 50 MHz and 300 MHz, networking demands randomly distributed between 5 Mbps to 50 Mbps, and storage demands randomly distributed between 10 MB to 400 MB. The requests need to be processed using a software package from a pool of 10 software packages of equal popularity. The central cloud and the fixed fog node are assumed to be preloaded with all software packages. We examine 1-10 preloaded software packages in each vehicle in the VF.

Five scenarios are considered. In scenario 1 (Cloud), all requests are assigned to the centralized cloud. This acts as a baseline. In scenario 2 (Cloud-Fog), the requests are assigned to the centralized cloud or the fixed fog node. In scenario 3 (Cloud-Fog-VF1), each request is assigned to either the cloud, the fixed fog node or a VF that has 5 vehicles. In scenario 4 (Cloud-Fog-VF2), each request is assigned to the cloud, the fixed fog node or a VF that has 10 vehicles. In scenario 5 (Cloud-Fog-VF3), each request is assigned to either the cloud, the fixed fog node or a VF that has 20 vehicles. Each scenario is evaluated with the two software deployments.

*Table I. Input Data for network devices*

| Network components | Capacity | Power consumption | Network components | Capacity | Power consumption |
|---|---|---|---|---|---|
| Vehicle processor | 240 MHz | 3.1 W [42] | Fog networking | 2.4 Gbps | 48 W [24] |
| Vehicle storage | 8 GB | 0.5 W [42] | Ethernet switch | 100 Gbps | 63.2 kW [30] |
| Vehicle WiFi | 54 Mbps | 0.207 W [30] | Edge router | 200 Gbps | 4.2 kW [30] |
| AP | 1.75 Gbps | 7.42 W [30] | Core router | 640 Gbps | 10.9 kW [30] |
| RSU | 27 Mbps | 7 W [30] | Cloud server | 4 GHz | 300 W [44] |
| ONU | 2.488 Gbps | 5 W [30] | Cloud storage | 75.6 TB | 4.9 kW [15] |
| OLT | 320 Gbps | 400 W [30] | Cloud switch | 320 Gbps | 3.8 kW [15] |
| Fog server | 2.7 GHz | 64.5 W [43] | Cloud router | 660 Gbps | 5.1 kW [15] |
| Fog storage | 120 GB | 10.5 W [43] | | | |

Figure 2 and Figure 3 show the total power consumption of the fives scenarios (Cloud, Cloud-Fog, Cloud-Fog-VF1, Cloud-Fog-VF2 and Cloud-Fog-VF3) deploying the same-type and random-type software packages in VF, respectively. The request allocations of the scenarios considered are shown in Figure 4 and Figure 5.

Figure 2 shows that the Cloud-Fog-VF1 and Cloud-Fog-VF2 power consumption decrease as the number of software packages available in the VF increases considering the same-type software deployment. This is because the availability of more software packages in the VF (the most efficient processing layer) allows it to process more requests and therefore reduces the power consumption. Preloading 6 software packages in the 5 vehicles of the Cloud-Fog-VF1 scenario and 8 software packages in the 10 vehicles of the Cloud-Fog-VF2 scenario allow the vehicles capacity to be fully utilized as seen in Figure 4. The requests beyond the capacity of the software availability of the VF are processed in the fixed fog and the cloud as seen in Figure 4. The power consumption of Cloud-Fog-VF3 scenario continually decreases as more software packages become available in the VF. This is because of the larger processing capacity of this scenario. Preloading the 10 software packages in each of the 20 vehicles results in fully utilising all of the processing capacity of the vehicles as seen Figure 4. Compared to the cloud scenario, the Cloud-Fog-VF1, Cloud-Fog-VF2, Cloud-Fog-VF3 scenarios achieved average power savings of 29%, 35% and 39% respectively with the same-type software deployment. Compared to the Cloud-Fog scenario, the corresponding power savings are 10%, 16% and 22% respectively.

With the random-type software deployment, deploying 5 software packages and 4 software packages, respectively, in each vehicle of Cloud-Fog-VF1 scenario and Cloud-Fog-VF2 scenario results in the minimum power consumption for the two scenarios, as seen in Figure 3. Deploying more software packages does not result in further reduction in power consumption as the preloaded software packages are enough to fully utilise the capacity of the VF as seen in Figure 5. For the Cloud-Fog-VF3 scenario, 3 software packages are needed to fully utilise the VF capacity. Compared to the Cloud scenario, the Cloud-Fog-VF1, Cloud-Fog-VF2, Cloud-Fog-VF3 scenarios achieve average power savings of 30%, 37% and 48% respectively. The corresponding power savings are 11%, 20% and 34% respectively compared with the Cloud-Fog scenario. Compared to the same-type software packages deployment, deploying random-type software packages results in up to 9% further power saving.

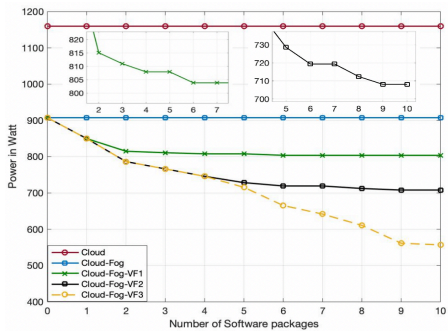

*Figure 2. Power consumption with same-type software deployment*

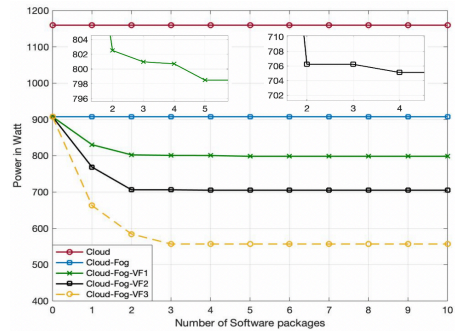

*Figure 3. Power consumption with random-type software deployment*

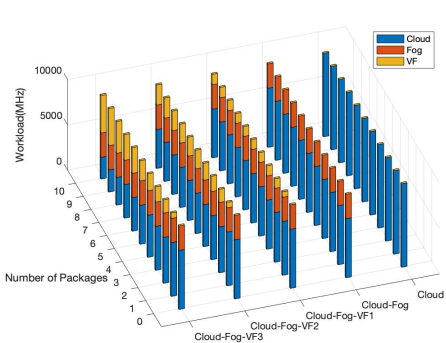

*Figure 4. Request allocation with same-type software deployment*

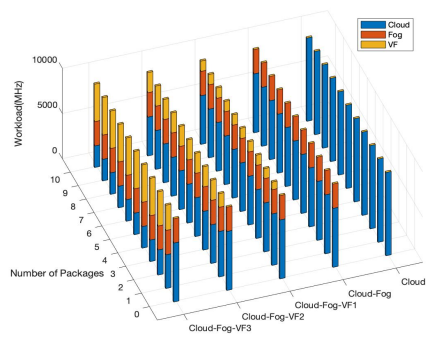

*Figure 5. Request allocation with random-type software deployment*

## 4. CONCLUSIONS

In this paper, we introduced an architecture that integrates a VF made up of vehicles clustered in a parking lot with a fixed fog node and the central cloud. We studied the problem of software matching in VF to improve the energy efficiency of this architecture. A MILP model is developed to minimize the overall power consumption considering two software packages deployment approaches. The results show that preloading the vehicles with the optimum number of software packages of the same type can result in power savings of up to 39% compared to processing in the cloud. Further power savings of 9% can be achieved by preloading software packages of random types.

ACKNOWLEDGEMENTS

The authors would like to acknowledge funding from the Engineering and Physical Sciences Research Council (EPSRC), through INTERNET (EP/H040536/1), STAR (EP/K016873/1) and TOWS (EP/S016570/1) projects. All data is provided in the results section of this paper.